\documentclass[12pt]{article}
\begin{document}

\begin{center}
\textbf{\Large \noindent On the origin of superselection rules and different 
solutions of Thirring model} \\ [0.5cm]
S.~E.~Korenblit, V.~V.~Semenov \\ [0.5cm]
\end{center}
\begin{tabular}{l}
\textit{\indent Irkutsk State University, 664003, Irkusk, Russia}\\
\textit{\indent e-mail: korenb@ic.isu.ru}
\end{tabular}
\begin{abstract}
The normal forms of different one- and two- parametric solutions of Thirring model are 
connected with each other by making use of generalized conformal shift transformations. 
A new alternative sources of superselection rules are shown and the ways of 
spontaneous symmetry breaking are discussed\footnote{This is extended 
version of manuscript of lectures presented by the author at Baikal Summer School on 
Physics of Elementary Particles and Astrophysics in 3-11 July 2011. 
The lectures are intended mainly for students and young researchers as an introductory 
course of QFT in two dimensions.}.
\end{abstract}
\begin{tabular}{l}
\hbox{PACS: 03.70.+k, 11.10.-z, 11.10.Kk, 11.15.Tk }  \hfill \\
\end{tabular}

\section{Introduction}
We have shown in the recent works \cite{ks_tt} that Thirring model is exactly solvable 
\cite{klaib,d_f_z,fab-iva} in fact due to intrinsic hidden exact linearizability of its 
Heisenberg equation (HEq) $2\partial_\xi \Psi_\xi(x)=-igJ^{-\xi}_{(\Psi)}(x)\Psi_\xi(x)$, 
and that the bosonization rules \cite{man} can acquire ``almost'' an operator sense only 
among the free fields operators $\chi(x),J_{(\chi)}(x)$ with unambiguously defined 
normal ordering procedure only for $Z_{(\chi)}(a)=1$. For Heisenberg currents with 
current's -- field's renormalization constant 
$Z_{(\Psi)}(a)=(-\Lambda^2 a^2)^{-{\overline{\beta}^2}/{4\pi}}$ 
these rules take place only in a weak sense:
\begin{eqnarray}
&&\!\!\!\!\!\!\!\!\!\!\!\!\!\!\!\!\!
\widehat{J}_{(\Psi)}^\mu (x)\stackrel{\rm w}{=}
\frac{\beta}{2\sqrt{\pi}}\,\widehat{J}_{(\chi)}^\mu (x)\stackrel{\rm w}{=}
-\,\frac{\beta}{2\pi}\,\epsilon^{\mu\nu}\partial_\nu\phi (x)=
\frac{\beta}{2\pi}\,\partial^\mu\varphi (x),  \mbox{ for:} 
\label{K_Za} \\
&&\!\!\!\!\!\!\!\!\!\!\!\!\!\!\!\!\!
\widehat{J}^0_{(\Psi)}(x)=
\lim\limits_{\widetilde{\varepsilon} \rightarrow 0}
\widehat{J}^0_{(\Psi)}(x;\widetilde{\varepsilon}),
\quad 
\widehat{J}^1_{(\Psi)}(x)=\lim\limits_{\varepsilon \rightarrow 0}
\widehat{J}^1_{(\Psi)}(x;\varepsilon),\; \mbox{ with:}
\label{bos-111} \\
&&\!\!\!\!\!\!\!\!\!\!\!\!\!\!\!\!\!
\widehat{J}^\nu_{(\Psi)}(x;a)=
Z^{-1}_{(\Psi)}(a)\left[\overline{\Psi}(x + a)\gamma^\nu \Psi (x)-
\langle 0|\overline{\Psi}(x + a)\gamma^\nu\Psi(x)|0\rangle\right],
\label{K_Z}
\end{eqnarray} 
and $\widetilde{\varepsilon}^\mu=-\epsilon^{\mu\nu}\varepsilon_\nu$, for the first  
$\widetilde{\varepsilon}^0 = \varepsilon^1\rightarrow 0$, but fixed  
$\widetilde{\varepsilon}^1 =\varepsilon^0$, $\varepsilon^2=-\widetilde{\varepsilon}^2>0$. 
The following variant of Oksak solution \cite{blot,oksak} of this model was obtained
\cite{ks_tt}:  
\begin{eqnarray}
&&\!\!\!\!\!\!\!\!\!\!\!\!\!\!\!\!\!\!\!\!
\Psi^{Ok}_\xi(x)={\cal N}_\varphi\left\{
\exp\left(-i2\sqrt{\pi}\left[{\varrho}_\eta^{-\xi}(x)+
\frac{\xi}{4}{\rm W}_\eta^\xi\right]\right)\right\}v_\xi ,\; \mbox{ with: }\; \xi=\pm,
\label{Psi_varho} \\
&&\!\!\!\!\!\!\!\!\!\!\!\!\!\!\!\!\!\!\!\!
2\sqrt{\pi}\,{\varrho}_\eta^{-\xi}(x)=
\overline{\alpha}\varphi^{-\xi}\left(x^{-\xi}\right)+
\overline{\beta}\varphi^{\xi}\left(x^\xi\right),\quad 
2\sqrt{\pi}\,{\rm W}_\eta^\xi=\overline{\alpha} Q^\xi-\overline{\beta}Q^{-\xi}, 
\label{varho_W} \\
&&\!\!\!\!\!\!\!\!\!\!\!\!\!\!\!\!\!\!\!\!
v_\xi =\widehat{\rm v}_\xi 
\exp\left\{-a_0\frac{\pi}{8}\cosh 2\eta\right\}, 
\quad 
\widehat{\rm v}_\xi=\left(\frac{\overline{\mu}}{2\pi}\right)^{1/2}
\left(\frac{\overline{\mu}}{\Lambda}\right)^{\overline{\beta}^2/{4\pi}}
e^{i\varpi-i\xi\Theta/4},
\label{M_2} \\
&&\!\!\!\!\!\!\!\!\!\!\!\!\!\!\!\!\!\!\!\!
\overline{\alpha}^2-\overline{\beta}^2= 4\pi, \quad
\overline{\beta}=\frac{\beta g}{2\pi}, \quad 
e^\eta=\frac{2\sqrt{\pi}}{\beta}=\sqrt{1+\frac{g}{\pi}}, \quad 
\overline{\alpha}\pm\overline{\beta}=2\sqrt{\pi}e^{\pm\eta},
\label{nweyi32}
\end{eqnarray}
defined for $x^\xi=x^0+\xi x^1$ in the space of massless pseudoscalar field 
$\phi(x)$: ${\cal P}c(k^1){\cal P}^{-1}=-c(-k^1)$, 
$[c(k^1),c^\dagger(q^1)]=4\pi k^0\delta(k^1-q^1)$, $c(k^1)|0\rangle=0$, for the scalar
charge $O=\phi(-\infty,x^0)-\phi(\infty,x^0)$ and pseudoscalar charge 
$O_5=\varphi(-\infty,x^0)-\varphi(\infty,x^0)$, and left $(\xi=+)$, right $(\xi=-)$ 
moves: $2\varphi^{\xi}\left(x^\xi\right)=\varphi(x)-\xi\phi(x)$ and charges 
$2Q^\xi=O-\xi O_5$, with: $\varepsilon(s)={\rm sgn}(s)$, 
\begin{eqnarray}
&&\!\!\!\!\!\!\!\!\!\!\!\!\!\!\!\!\!\!
k^0=|k^1|, \quad {\rm P}\frac 1{k^1}=\frac{\varepsilon\left(k^1\right)}{|k^1|},
\quad \frac 14\left({\rm P}\frac 1{k^1}-\frac{\xi}{k^0}\right)=
\frac{-\xi\theta\left(-\xi k^1\right)}{2 k^0}, \;\mbox{ and:}  
\label{P_e_k0} \\
&&\!\!\!\!\!\!\!\!\!\!\!\!\!\!\!\!\!\!
\phi(x) = \frac{1}{2\pi} \int\limits_{-\infty}^\infty\frac{d k^1}{2 k^0}
\left[c\left(k^1\right)e^{-i(kx)}+c^\dagger\left(k^1\right)e^{i(kx)}\right]\equiv
\phi^{(+)}(x)+\phi^{(-)}(x),
\label{phi_+_-} \\
&&\!\!\!\!\!\!\!\!\!\!\!\!\!\!\!\!\!\!
\varphi(x) = \frac{1}{2\pi}\int\limits_{-\infty}^\infty d k^1\,{\rm P}\frac 1{k^1} 
\left[c\left(k^1\right)e^{-i(kx)}+c^\dagger\left(k^1\right)
e^{i(kx)}\right],
\label{varphi_+_-} \\
&&\!\!\!\!\!\!\!\!\!\!\!\!\!\!\!\!\!\!
\varphi^{\xi(+)}(s)= 
-\,\frac{\xi}{2\pi}\int\limits_{-\infty}^\infty \frac{d k^1}{2 k^0}
\theta\left(-\xi k^1\right)c(k^1) e^{-i k^0 s}, \quad 
\varphi^{\xi(-)}(s)=\left[\varphi^{\xi(+)}(s)\right]^\dagger,
\label{ph_pm} \\
&&\!\!\!\!\!\!\!\!\!\!\!\!\!\!\!\!\!\!
Q^{\xi(+)}(\widehat{x}^0)=\lim_{L\rightarrow\infty} \frac{i\xi}{2}
\int\limits_{-\infty}^\infty d k^1 \theta\left(-\xi k^1\right) c(k^1) 
e^{-ik^0 \widehat{x}^0}\delta_L (k^1), \quad Q^{\xi(-)}=\left[Q^{\xi(+)}\right]^\dagger,
\label{Q_pm} \\
&&\!\!\!\!\!\!\!\!\!\!\!\!\!\!\!\!\!\!
\left[\varphi(x), O\right]=\left[\phi(x), O_5\right]=i,\quad
2i O_5=c(0)-c^\dagger(0),  
\label{Q_mp} \\
&&\!\!\!\!\!\!\!\!\!\!\!\!\!\!\!\!\!\!
\left[\varphi^\xi\left(s\right),\varphi^{\xi'}\left(\tau\right)\right]=
-\frac{i}{4}\varepsilon(s - \tau)\delta_{\xi, \xi'}, \quad
\left[\varphi^\xi(s), Q^{\xi'}\right]=\frac{i}{2}\delta_{\xi,\xi'},
\label{K_9} \\
&&\!\!\!\!\!\!\!\!\!\!\!\!\!\!\!\!\!\!
\left[\varphi^{\xi(\pm)}(s),\varphi^{\xi'(\mp)}(\tau)\right]=
\mp\frac{\delta_{\xi,\xi'}}
{4\pi}\ln\biggl(i\overline{\mu}\Bigl\{\pm(s-\tau)-i0\Bigr\}\biggr),
\label{nblaie16} \\
&&\!\!\!\!\!\!\!\!\!\!\!\!\!\!\!\!\!\!
\left[\varphi^{\xi(\pm)}(s),  Q^{\xi'(\mp)}\right]=
\frac{i}{4}\delta_{\xi, \xi'}, \quad
\left[ Q^{\xi(\pm)}, Q^{\xi'(\mp)}\right]=\pm a_0 \delta_{\xi, \xi'}, 
\label{nblaie19} \\
&&\!\!\!\!\!\!\!\!\!\!\!\!\!\!\!\!\!\!
a_0=a_0(L)=\pi\int\limits_{0}^\infty d k^1 k^0\left(\delta_L(k^1)\right)^2, \quad 
L\rightarrow\infty, \quad \lim_{L\rightarrow\infty}\delta_L(k^1)=\delta(k^1). 
\label{D_a0} 
\end{eqnarray}
For the Eq. (\ref{Psi_varho}) constructed in \cite{ks_tt} as dynamical mapping (DM), 
${\cal N}_\varphi$ means the normal ordering with respect to $c(k^1)$, and the 
ultraviolet cut-off $\Lambda$ \cite{fab-iva} is introduced for the field regularization 
(\ref{M_2}). 
The volume cut-off function $\delta_L(k^1)$ was invented for the charge regularization 
(\ref{Q_pm}), leading to nonnegative constant $a_0$ (\ref{D_a0}) in the commutator 
(\ref{nblaie19}), connected with the structure of vacuum state of chosen field 
representation \cite{d_f_z,fab-iva,blot,oksak,mps_2,ks_tt}. 
Unfortunately any of such (continuous or not) cut-off regularization induces 
the fictitious and non-physical $\widehat{x}^0$ - dependence of formally conserved 
charges $O$, $O_5$ or $Q^\xi$ and their frequency parts (\ref{Q_pm}), and, in general, 
destroys the above topological definitions of these charges. 
Such kind of $\widehat{x}^0$ - dependence, being an artifact of the charge's 
regularization (\ref{Q_pm}), should be eliminated at the end of calculation. 

The observed weak linearization of HEq together with the nonlinearity of DM 
(\ref{Psi_varho}) and the applied in a weak sense initial conditions at $x^0=0$, 
have allowed \cite{ks_tt} to overcome the restrictions of Haag theorem, by removing the 
problems again into the representation construction of physical fields: at first as 
reducible massless free Dirac fields $\chi(x)$ corresponding to 
$\overline{\beta}=\eta=g=0$ in above formulas (\ref{K_Za})--(\ref{nweyi32}), 
and then as irreducible massless (pseudo) scalar fields $(\phi(x)), \varphi(x)$  
(\ref{P_e_k0})--(\ref{D_a0}), taken here mutually dual and coupled by the relations 
($\varepsilon(x^1)={\rm sgn}(x^1)$):
\begin{eqnarray}
&&\!\!\!\!\!\!\!\!\!\!\!\!\!\!\!\!\!\!
\left. 
\begin{array}{c} \phi(x) \\ \varphi(x) \end{array}
\right\}
=-\frac{1}{2}\int\limits_{-\infty}^\infty dy^1
\varepsilon \left(x^1-y^1\right)\partial_0 \left\{
\begin{array}{c}
\varphi\left(y^1,x^0\right). \\ \phi\left(y^1,x^0\right). \end{array} \right. 
\label{K_5}
\end{eqnarray}
These free fields arisen \cite{ks_tt} as Schr\"odinger physical fields, in fact play 
a role of asymptotic ones \cite{blot}. Due to automatical elimination of zero mode's 
contributions for (\ref{varphi_+_-}) and for generating functional of pseudo scalar 
field \cite{fab-iva}, the 
representation space \cite{ks_tt} chosen here relaxes the problem of non-positivity of 
its inner product induced by Wightman function (\ref{nblaie16}) \cite{blot,oksak,mps_2}. 

\section{Other solutions and superselection rules}

Now we wish to connect the Oksak solution (\ref{Psi_varho})--(\ref{M_2}) with another 
known solutions of Thirring model \cite{man,mps_2}. To this end we use the 
properly unitary transformation of conformal 
shift \cite{oksak} for the fields $\varphi^{\xi}$, generalized in the following way. 
By making use of the relations (\ref{Q_mp})--(\ref{nblaie19}), we consider the family 
of solutions $\Psi(x,{\sigma})={\rm K}^{-1}_\sigma\Psi^{Ok}(x){\rm K}_\sigma$, marked by 
arbitrary real parameter $\sigma$: 
\begin{eqnarray}
&&\!\!\!\!\!\!\!\!\!\!\!\!\!\!\!\!\!\!\!\!
{\rm K}_\sigma=\exp{\rm X}_\sigma,\quad 
{\rm X}_\sigma=i\sigma\frac{\overline{\xi}}4
\left(Q^{-\overline{\xi}}Q^{-\overline{\xi}}-Q^{\overline{\xi}}Q^{\overline{\xi}}\right)
=i\frac \sigma 4 OO_5, \quad \overline{\xi},\xi=\pm 
\label{K_X_s} \\
&&\!\!\!\!\!\!\!\!\!\!\!\!\!\!\!\!\!\!\!\!
\Psi_\xi(x,{\sigma})=
{\rm K}^{-1}_\sigma\Psi^{Ok}_\xi(x){\rm K}_\sigma=
{\cal N}_\varphi\left\{e^{R_\xi(x,\sigma)}\right\} v_{\xi}(\sigma),
\label{K_Psi_s} \\
&&\!\!\!\!\!\!\!\!\!\!\!\!\!\!\!\!\!\!\!\!
R_\xi(x,\sigma)=
-i\left[2\sqrt{\pi}\,{\varrho}_\eta^{-\xi}(x)
+\frac{\xi}{4}(\overline{\alpha}+\sigma\overline{\beta}) Q^{\xi}
-\frac{\xi}{4}(\overline{\beta}+\sigma\overline{\alpha})Q^{-\xi}\right],
\label{R_Psi_s} \\
&&\!\!\!\!\!\!\!\!\!\!\!\!\!\!\!\!\!\!\!\!
v_\xi(\sigma)=v_\xi 
\exp\left\{-a_0\frac{\pi}{8}\left[\sigma^2\cosh 2\eta+
2\sigma\sinh 2\eta\right]\right\}.
\label{v_xi_s} 
\end{eqnarray}
For arbitrary $\sigma$ this solution obeys the same canonical anticommutation relations 
(CAR) and the same 
bosonization rule (\ref{K_Za})--(\ref{K_Z}) with the same renormalization constant 
$Z_{(\Psi)}(a)$. The parameter $a_0$ may be adsorbed into the regularization parameter 
$\overline{\mu}$ by the re-scaling substitution, which unlike the Oksak (\ref{M_2})
and free cases, now depends on Thirring coupling constant $g$ (\ref{nweyi32}): 
\begin{eqnarray}
\overline{\mu}\longmapsto\overline{\overline{\mu}}\,
\exp\left\{a_0\frac \pi 4\left(\sigma^2+1+2\sigma\tanh 2\eta\right)\right\}.  
\label{mu_a0eta}
\end{eqnarray}
By using (\ref{K_5}) and above topological definitions both of charges $O,O_5$, it is a 
simple matter to 
check that $\sigma=\pm 1$ gives the two types of Mandelstam solution \cite{man}, while  
$\sigma=-\coth 2\eta$ corresponds to normal form of solution of Morchio et al. 
\cite{mps_2}. This again demonstrates the advantages of normal ordered form of DM:
\begin{eqnarray}
&&\!\!\!\!\!\!\!\!\!\!\!\!\!\!\!\!\!\!\!\!
\Psi_\xi(x,1)=
{\cal N}_\varphi\left\{e^{R_\xi(x,1)}\right\} \widehat{\rm v}_\xi 
\exp\left\{-a_0\frac{\pi}{4}e^{2\eta}\right\}, \quad \sigma=1,
\label{Man_+1} \\
&&\!\!\!\!\!\!\!\!\!\!\!\!\!\!\!\!\!\!\!\!
R_\xi(x,1)=-i\sqrt{\pi}\left[\xi e^{-\eta}\phi(x)-e^{\eta}
\int\limits^{x^1}_{-\infty}dy^1\,\partial_0\phi(y^1,x^0)\right],
\label{Man_+1R} \\ 
&&\!\!\!\!\!\!\!\!\!\!\!\!\!\!\!\!\!\!\!\!
\Psi_\xi(x,-1)=
{\cal N}_\varphi\left\{e^{R_\xi(x,-1)}\right\}\widehat{\rm v}_\xi 
\exp\left\{-a_0\frac{\pi}{4}e^{-2\eta}\right\}, \quad \sigma=-1, 
\label{Man_-1} \\
&&\!\!\!\!\!\!\!\!\!\!\!\!\!\!\!\!\!\!\!\!
R_\xi(x,-1)=-i\sqrt{\pi}\left[e^{\eta} \varphi(x)+
\xi e^{-\eta}\int\limits_{x^1}^{\infty}dy^1\,\partial_0\varphi(y^1,x^0)\right],
\label{Man_-1R} \\
&&\!\!\!\!\!\!\!\!\!\!\!\!\!\!\!\!\!\!\!\!
\Psi_\xi(x,-\coth 2\eta)={\cal N}_\varphi\left\{e^{R_\xi(x,-\coth 2\eta)}\right\}
v_\xi( \sigma=-\coth 2\eta),
\label{MPS} \\
&&\!\!\!\!\!\!\!\!\!\!\!\!\!\!\!\!\!\!\!\!
R_\xi(x,-\coth 2\eta)=-i\left[2\sqrt{\pi}\,{\varrho}_\eta^{-\xi}(x)
+\xi \frac{\pi}{2}\left(\frac{ Q^{\xi}}{\overline{\alpha}}+
\frac{Q^{-\xi}}{\overline{\beta}}\right)\right],
\label{MPS_R} \\
&&\!\!\!\!\!\!\!\!\!\!\!\!\!\!\!\!\!\!\!\!
v_\xi(\sigma=-\coth 2\eta)=\widehat{\rm v}_\xi 
\exp\left\{-a_0\frac\pi 8\frac{\cosh 2\eta}{\sinh^2 2\eta}\right\}. 
\label{MPS_V} 
\end{eqnarray} 
We would like to point out that $\sigma=1$ corresponds to DM (\ref{Man_+1}), 
(\ref{Man_+1R}) ``onto'' the pseudoscalar field $\phi(x)$, while $\sigma=-1$ gives 
another form (\ref{Man_-1}), (\ref{Man_-1R}) of  Mandelstam solution with bosonization 
``onto'' the scalar field $\varphi(x)$, and that unlike (\ref{MPS})--(\ref{MPS_R}), the 
original solution of Morchio et al. \cite{mps_2} has $a_0=0$, as well as the original 
Oksak solution \cite{blot,oksak}, but contains all Klein factors outside the normal form, 
so its renormalization constant remains to be unknown. 
We have used here that $2{\varrho}_\eta^{-\xi}(x)=\varphi(x)e^\eta+\xi\phi(x)e^{-\eta}$, 
$2{\rm W}^{-\xi}_{\eta}=O e^{-\eta}+\xi O_5e^\eta$, and have utilized for brevity both 
the definitions of last identity (\ref{nweyi32}) read also as: 
$\overline{\alpha}+\overline{\beta}=4\pi/\beta=2\sqrt{\pi} e^\eta$, 
$\overline{\alpha}-\overline{\beta}=\beta=2\sqrt{\pi} e^{-\eta}$.  

There is another important maybe improperly unitary transformation of the solutions 
(\ref{K_Psi_s}), which introduces the two-parametric extension of Oksak solution 
(\ref{Psi_varho}) and for arbitrary $\sigma, \rho$ obeys again the same CAR and the 
bosonization rule (\ref{K_Za}) with the same renormalization constant $Z_{(\Psi)}(a)$:
\begin{eqnarray}
&&\!\!\!\!\!\!\!\!\!\!\!\!\!\!\!\!\!\!\!\!
\Psi(x,\sigma,\rho)={\cal L}_{\rho}^{-1}\Psi(x,\sigma){\cal L}_{\rho}=
{\rm K}^{-1}_\sigma\Psi(x,0,\rho){\rm K}_\sigma,\;\mbox{and for }\;
\overline{\xi},\xi=\pm : 
\label{K_cL} \\
&&\!\!\!\!\!\!\!\!\!\!\!\!\!\!\!\!\!\!\!\!
{\cal L}_{\rho}=\exp{\rm Y}_\rho,\quad
{\rm Y}_\rho=-\frac i2 \rho\,Q^{\overline{\xi}}Q^{-\overline{\xi}}=
-\,\frac i8 \rho\left(O^2-O^2_5\right), 
\label{cL_rho} \\
&&\!\!\!\!\!\!\!\!\!\!\!\!\!\!\!\!\!\!\!\!
\Psi_\xi (x,\sigma,\rho)=
{\cal L}_{\rho}^{-1}\Psi_\xi (x,\sigma){\cal L}_{\rho}=
{\cal N}_\varphi\left\{e^{R_\xi(x,\sigma,\rho)}\right\} v_{\xi}(\sigma,\rho),
\label{psi_rho_sigm} \\
&&\!\!\!\!\!\!\!\!\!\!\!\!\!\!\!\!\!\!\!\!
R_\xi(x,\sigma,\rho)=
-i2\sqrt{\pi}\left[{\varrho}_\eta^{-\xi}(x)
+\frac{\Sigma^\xi_+}{8}Q^{\xi}+\frac{\Sigma^\xi_-}{8}Q^{-\xi}\right],
\label{R_rh_Sg_Q} \\
&&\!\!\!\!\!\!\!\!\!\!\!\!\!\!\!\!\!\!\!\!
v_\xi(\sigma,\rho)= \widehat{\rm v}_\xi 
\exp\left\{-a_0\frac{\pi}{32}\left[\left(\Sigma^\xi_-\right)^2+
\left(\Sigma^\xi_+\right)^2\right]\right\}, \;\mbox{ with:}
\label{v_xi_r_S} \\
&&\!\!\!\!\!\!\!\!\!\!\!\!\!\!\!\!\!\!\!\!
\Sigma^\xi_{\pm}=e^{-\eta}\left[\xi(1-\sigma)+\rho\right]\pm 
e^{\eta}\left[\xi(1+\sigma)+\rho\right]. 
\label{sgm_Sgm_01} 
\end{eqnarray}
For divergent value of $a_0$ at $L\to\infty$, e.g. for usual box \cite{ks_tt}, the $\xi$- 
dependence of the last exponential of c - number spinor (\ref{v_xi_r_S}), 
(\ref{sgm_Sgm_01}) leads to non-physical in general non-renormalizable infrared 
divergences of every components of the field (\ref{psi_rho_sigm}). For arbitrary 
$\rho$ this $\xi$- dependence eliminates only for the generalized solution (\ref{MPS}) 
with $\sigma=-\coth 2\eta$, that transcribes (\ref{psi_rho_sigm})--(\ref{sgm_Sgm_01}) as: 
\begin{eqnarray}
&&\!\!\!\!\!\!\!\!\!\!\!\!\!\!\!\!\!\!\!\!
R_\xi(x,-\coth 2\eta,\rho)= 
-i\left[2\sqrt{\pi}\,{\varrho}_\eta^{-\xi}(x)
+\xi\frac{\pi}{2}\left(\frac{Q^{\xi}}{\overline{\alpha}}+
\frac{Q^{-\xi}}{\overline{\beta}}\right)+\rho\frac{\sqrt{\pi}}2{\rm W}_\eta^{\xi}\right], 
\label{Mor_R_r} \\
&&\!\!\!\!\!\!\!\!\!\!\!\!\!\!\!\!\!\!\!\!
v_\xi(-\coth 2\eta,\rho)=\widehat{\rm v}_\xi 
\exp\left\{-a_0\frac{\pi}{8}\cosh 2\eta
\left[\frac{1}{\sinh^2 2\eta}+\rho^2\right]\right\}. 
\label{v_Mor_r} 
\end{eqnarray}

Let us turn to the vacuum expectation value (VEV) of the strings of these fields 
(\ref{psi_rho_sigm}). Following \cite{blot} it is enough to consider only the product: 
\begin{eqnarray}
&&\!\!\!\!\!\!\!\!\!\!\!\!\!\!\!\!\!\!\!\!
\left\langle 0\left|\prod\limits_{{\rm i}=1}^p
\Psi_{\xi_{\rm i}}^{(l_{\rm i})}(x_{\rm i},\sigma.\rho)
\right|0 \right\rangle,\;\mbox{ with: }\; l_{\rm i}=+1,\;\mbox{ for }\;\Psi_{\rm i},
\;\mbox{ and: }\; l_{\rm i}=-1, \;\mbox{ for }\; \Psi^\dagger_{\rm i}.
\label{Psi_ppp} 
\end{eqnarray}
By virtue of the known relations for any numbers $\gamma_{\rm i}$, with 
${\rm i,k}=1\div p$, and for any linearly depended on $c(k^1),c^\dagger(k^1)$ operators 
$l_{\rm i}R_{\xi_{\rm i}}(x_{\rm i},\sigma,\rho)\mapsto 
{\cal R}_{\rm i}={\cal R}_{\rm i}^{(+)}+{\cal R}_{\rm i}^{(-)}$: 
\begin{eqnarray}
&&\!\!\!\!\!\!\!\!\!\!\!\!\!\!\!\!\!\!\!\!
\sum\limits_{{\rm i}<{\rm k}}^p\gamma_{\rm i}\gamma_{\rm k}=
\frac{1}{2}\sum\limits_{{\rm i}\neq{\rm k}}^p\gamma_{\rm i}\gamma_{\rm k}
=\frac{1}{2}\left(\sum\limits_{{\rm i}=1}^p \gamma_{\rm i}\right)^2
- \frac{1}{2}\sum\limits_{{\rm i}=1}^p \gamma_{\rm i}^2,
\label{LL_LL}  \\
&&\!\!\!\!\!\!\!\!\!\!\!\!\!\!\!\!\!\!\!\!
\prod\limits_{{\rm i}=1}^p {\cal N}\Bigl\{\exp\left[{{\cal R}_{\rm i}}\right]\Bigr\}= 
\exp\left\{\sum\limits_{{\rm i}<{\rm k}}^p
\left[{\cal R}_{\rm i}^{(+)},{\cal R}_{\rm k}^{(-)}\right]\right\}
{\cal N} \left\{\exp\left(\sum\limits_{j=1}^p {\cal R}_j\right)\right\},
\label{AAA_BBB} 
\end{eqnarray}
the corresponding VEV of the string of the fields 
(\ref{psi_rho_sigm})--(\ref{sgm_Sgm_01}) reads\footnote{Here for 
$\xi_{\rm i}\pm \xi_{\rm k}$ and $\xi_{\rm i}\xi_{\rm k}$,
we use $\xi_{\rm i},\xi_{\rm k}=+1,-1$, and: 
$\delta_{\xi_{\rm i},\xi_{\rm k}}=(1+\xi_{\rm i}\xi_{\rm k})/2$, 
$\delta_{\xi_{\rm i},\pm 1}=(1\pm \xi_{\rm i})/2$.}:
\begin{eqnarray}
&&\!\!\!\!\!\!\!\!\!\!\!\!\!\!\!\!\!\!
\left(\Lambda^{\overline{\beta}^2/4\pi}\sqrt{2\pi}\right)^p
\left\langle 0\left|\prod\limits_{{\rm i}=1}^p
\Psi_{\xi_{\rm i}}^{(l_{\rm i})}(x_{\rm i},\sigma,\rho)\right|0 \right\rangle=
\left\langle 0\left|{\cal N}_\varphi\left\{\exp\left(\sum\limits_{j=1}^p
{\cal R}_j\right)\right\}\right|0\right\rangle  
\nonumber \\
&&\!\!\!\!\!\!\!\!\!\!\!\!\!\!\!\!\!\!
\cdot
\exp\left\{i\varpi{\cal S}_p-i\frac{\Theta}{4}{\cal S}_{p5}\right\}\,
\exp\left\{
\frac 14\left[e^{2\eta}{\cal S}^2_p+e^{-2\eta}{\cal S}^2_{p5}\right]\ln\overline{\mu}
\right\}
\nonumber \\
&&\!\!\!\!\!\!\!\!\!\!\!\!\!\!\!\!\!\!
\cdot 
\exp\left\{-a_0\frac{\pi}{16}\left(
e^{2\eta}\biggl[(1+\sigma){\cal S}_p+\rho{\cal S}_{p5}\biggr]^2+ 
e^{-2\eta}\biggl[(1-\sigma){\cal S}_{p5}+\rho{\cal S}_{p}\biggr]^2
\right)\right\}
\nonumber \\
&&\!\!\!\!\!\!\!\!\!\!\!\!\!\!\!\!\!\!
\cdot
\prod\limits_{{\rm i}<{\rm k}}^p \left\{e^{i \pi(\xi_{\rm i}-\xi_{\rm k})}
\left[\frac{x_{\rm i}^{-}-x_{\rm k}^{-}-i0}{x_{\rm i}^{+}-x_{\rm k}^{+}-i0}
\right]^{\xi_{\rm i}+\xi_{\rm k}} 
\left[i0\, \varepsilon (x_{\rm i}^{0}-x_{\rm k}^{0})-(x_{\rm i}-x_{\rm k})^2 \right]^
{e^{2\eta}+\xi_{\rm i}\xi_{\rm k} e^{-2\eta}}\right\}^{l_{\rm i} l_{\rm k}/4}.
\label{VV_rho}
\end{eqnarray}
This expression provides at least five independent sources of superselection rules 
\cite{klaib,blot} that usually are associated only with conservation of scalar field's 
(vector current's) charge $O$ and pseudoscalar field's (pseudovector current's) charge 
$O_5$, respectively. For the $p$ -point Wightman function (\ref{VV_rho}) there are:
\begin{eqnarray}
&&\!\!\!\!\!\!\!\!\!\!\!\!\!\!\!\!\!\!\!\!
{\cal S}_{p}\equiv\sum\limits_{{\rm i}=1}^p l_{\rm i} \Rightarrow 0, \qquad 
{\cal S}_{p5}\equiv \sum\limits_{{\rm i}=1}^p l_{\rm i} \xi_{\rm i} \Rightarrow 0. 
\label{ssr_2}
\end{eqnarray}   
The first one defined by Oksak and Morchio et al., due to above mentioned conservation 
both of charges, originates from the VEV of normal exponential in the r.h.s., taken, 
instead of $|0\rangle$, for the vacuum state $|\widehat{\upsilon}\rangle$ 
averaged with respect to the field-translation gauge group like (\ref{gauge}) below, 
leading to \cite{blot,oksak,mps_2}: 
\begin{eqnarray}
&&\!\!\!\!\!\!\!\!\!\!\!\!\!\!\!\!\!\!\!\!
\left\langle \widehat{\upsilon}\left|{\cal N}_\varphi\left\{\exp\left(\sum\limits_{j=1}^p
{\cal R}_j\right)\right\}\right|\widehat{\upsilon}\right\rangle\Longrightarrow 
\delta_{{\cal S}_{p},0}\,\delta_{{\cal S}_{p5},0}.
\label{ssr_3} 
\end{eqnarray}
For the usual non-degenerate vacuum state $|0\rangle$ this VEV is equal to 1 identically 
\cite{man}. Nevertheless, these rules (\ref{ssr_2}) arise again from the second line of 
(\ref{VV_rho}) at the limit $\overline{\mu}\to 0$ as a natural condition of nonzero 
result \cite{fab-iva}. 
We can suggest now three additional sources of these rules. The third one is the 
$\varpi$ - and $\Theta$ - independence condition for the VEV (\ref{Psi_ppp}), 
(\ref{VV_rho}), the fourth one follows from its independence on the regularization 
parameter $a_0$, and the fifth one follows from its independence on $\sigma$ and $\rho$
whenever the corresponding transformations (\ref{K_X_s}), (\ref{cL_rho}) leave the vacuum 
invariant. Obviously the $a_0$ - independence of VEV (\ref{VV_rho}) automatically means 
its $\sigma,\rho$ - independence and vice versa. 

The independence on the initial values of overall and relative phases has purely 
fermionic nature and does not reduce to the (pseudo) scalar field-translation gauge 
symmetry (\ref{gauge}), which can only shift their random initial values 
$\varpi$ and $\Theta$. 
It is worth to note the both superselection rules (\ref{ssr_2}) leave necessary 
for VEV (\ref{Psi_ppp}) only the ultraviolet renormalization. Whence, only the last line 
of (\ref{VV_rho}) with product over ${\rm i}<{\rm k}$ survives, which does not depend on 
the parameters $\overline{\mu}$, $\sigma$, $\rho$, $a_0$, $\varpi$, $\Theta $, and  
gives the well known expression for the $p$ -point functions \cite{klaib,man,blot} with 
correct dynamical dimensions.

The expression (\ref{VV_rho}) provides an interesting exchange symmetry between the 
superselection rules (\ref{ssr_2}) or between the respective gauge symmetries. 
Due to $\xi^2_{\rm i}=1$, $l^2_{\rm i}=1$, the substitutions: 
$l_{\rm i}\mapsto l_{\rm i}\xi_{\rm i}$, 
e.g. ${\cal S}_{p}\rightleftharpoons {\cal S}_{p5}$, with $\eta\mapsto -\eta$, 
$\sigma\mapsto -\sigma$, gives the same expression of VEV (\ref{VV_rho}) up to inessential 
overall phase factor equal to:  
\begin{eqnarray}
&&\!\!\!\!\!\!\!\!\!\!\!\!\!\!\!\!\!\!\!\!
\exp\left\{i\left({\cal S}_{p5}-{\cal S}_p\right)
\left[\varpi+\frac{\Theta}{4}\right]\right\}
\exp\left\{\{\pm\}i\frac{\pi}2{\cal S}_{p5}{\cal S}_{p}\right\}
\exp\left\{[\pm]i\frac{\pi}2\sum^p_{{\rm k}=1}\xi_{\rm k}\right\},
\label{factor}
\end{eqnarray}
which does not depend on the both sign arbitrarinesses here, because for any integer $N$: 
$e^{-i\pi N}=e^{i\pi N}$ as an overall multiplier, and for the even number $p$ the last 
sum also leads to even number. 

\section{Discussion and Conclusions}

Since $a_0$ (\ref{D_a0}) itself makes sense of regularization parameter, it should 
disappear in physical quantities. But for any real $\upsilon$ it defines the VEV of the 
operator of the field-translation gauge transformation \cite{blot,i_z}, e.g.: 
\begin{eqnarray}
&&\!\!\!\!\!\!\!\!\!\!\!\!\!\!\!\!\!\!
\exp\left\{i\upsilon O_5\right\}\phi(x)\exp\left\{-i\upsilon O_5\right\}=
\phi(x)+\upsilon, 
\label{gauge} \\
&&\!\!\!\!\!\!\!\!\!\!\!\!\!\!\!\!\!\!
\langle 0|\exp\left\{i\upsilon O_5\right\}|0\rangle\equiv\langle 0|\upsilon\rangle
=\exp\left\{-a_0\upsilon^2\right\}, 
\label{VEV_0}
\end{eqnarray}
which is well known in quantum theory of free massless (pseudo) scalar field \cite{i_z}.  
The state $|\upsilon\rangle$ is a coherent state of harmonic oscillator, which 
according to (\ref{Q_mp}) corresponds to zero mode $k^1=0$ for $L\to\infty$, and whenever 
simultaneously $a_0\to+\infty$, e.g., for the usual box of the length $2L$ \cite{ks_tt}, 
then $\langle 0|\upsilon\rangle\Rightarrow 0$ and the state $|\upsilon\rangle$ defines 
another orthogonal vacuum state of the degenerate family \cite{fab-iva,i_z}. 
The finite $a_0$ for any continuous piecewise-smooth function of volume 
cut-off regularization means the charge definition, which has nothing to do with 
previous thermodynamic limit and corresponds to another vacuum structure of the 
representation space of (pseudo) scalar field \cite{ks_tt}. 
A randomization of function $k^0\delta_L(k^1)$ and corresponding value of $a_0$ may be 
used for infrared ``stabilization'' of Wightman function (\ref{nblaie16}), which   
replaces the regularization parameter $\overline{\mu}$ to some 
finite scale $M$ \cite{fab-iva}. 

We see that, unlike Schwinger model \cite{blot,rub}, as long as we deal with the solutions 
(\ref{psi_rho_sigm})--(\ref{sgm_Sgm_01}) of above ``phase decoupled'' HEqs for $\xi=\pm$, 
that preserve the $\varpi$ and $\Theta$ arbitrariness, both of superselection rules 
(\ref{ssr_2}) and the conservation both of currents should be fulfilled independently of 
chosen phase of the theory, including the phase with spontaneously broken chiral symmetry 
\cite{fab-iva}. 

Only when both the rules (\ref{ssr_2}) are fulfilled the VEV (\ref{VV_rho}) does not 
depend on the regularization and transformation parameters: $\overline{\mu}$, 
$a_0$, and $\sigma$, $\rho$, $\varpi$, $\Theta$, and on the choice of volume cut-off 
regularization. Whereas the discarding of second superselection rule (\ref{ssr_2}) 
inevitably spoils the $\sigma$, $\rho$, and $\Theta $ -- invariance of this $p$ -point 
fermionic Wightman function and its independence of the parameters $\overline{\mu}$ and 
$a_0$. So, they should be fixed by some additional conditions \cite{fab-iva}, what seems  
impossible for the regularization dependent value of $a_0$. At the same time, the DM onto 
the free massive fields \cite{fab-iva,raja} will be free from that parameters. 

Formally from this view point the breaking of the second rule (\ref{ssr_2}) can be 
achieved or by introducing the mass term into HEq ``by hand'' \cite{fab-iva}, or 
otherwise, by excluding $a_0$ via taking Mandelstam solution with $\sigma=1$, $\rho=0$, 
supplemented with fixing of the values $\overline{\mu}\mapsto M$ and $\Theta$ 
\cite{fab-iva}. However, since one of the gauge symmetries remains unbroken: 
$O|0\rangle\Rightarrow 0$, all the solutions (\ref{K_Psi_s}), 
(\ref{Man_+1})--(\ref{MPS_V}) being connected by transformation (\ref{K_X_s}), refer 
of course to the same vacuum state $|0\rangle$ regardless the value of $\sigma$. 

In spite of the impossibility to remove the $\overline{\mu}$ and $a_0$ - dependence for 
$\rho\neq 0$ with any $\sigma$, whenever only the first of superselection roles 
(\ref{ssr_2}) is fulfilled, all these parameters can be adsorbed into the one parameter 
$M$ by the re-scaling substitution: 
\begin{eqnarray}
&&\!\!\!\!\!\!\!\!\!\!\!\!\!\!\!\!\!\!
\overline{\mu}\longmapsto 
M\exp\left\{a_0\frac{\pi}4\left[e^{4\eta}\rho^2+(1-\sigma)^2\right]\right\}, 
\label{mu_Ma0}
\end{eqnarray}
leading to the factor $M^{{\rm D}_p}$ with the same ``dynamical dimension'' 
${\rm D}_p=e^{-2\eta}{\cal S}^2_{p5}/4$ in (\ref{VV_rho}). 

The authors thank Y. Frishman, A.N. Vall, S.V. Lovtsov, V.M. Leviant, 
and participants of seminar in LTPh JINR for useful discussions.

This work was supported in part by the RFBR (project N 
09-02-00749) and by the program ``Development of Scientific 
Potential in Higher Schools'' (project N 2.2.1.1/1483, 
2.1.1/1539).



\begin{thebibliography}{99}
 \expandafter\ifx\csname natexlab\endcsname\relax\def\natexlab#1{#1}\fi
 \expandafter\ifx\csname bibnamefont\endcsname\relax
 \def\bibnamefont#1{#1}\fi
 \expandafter\ifx\csname bibfnamefont\endcsname\relax
 \def\bibfnamefont#1{#1}\fi
 \expandafter\ifx\csname citenamefont\endcsname\relax
 \def\citenamefont#1{#1}\fi
 \expandafter\ifx\csname url\endcsname\relax
 \def\url#1{\texttt{#1}}\fi
 \expandafter\ifx\csname urlprefix\endcsname\relax\def\urlprefix{URL }\fi
 \providecommand{\bibinfo}[2]{#2}
 \providecommand{\eprint}[2][]{\url{#2}}

{\footnotesize 

\bibitem {ks_tt}
  \emph{Korenblit~S.~E., Semenov~V.V.} Massless Thirring model in canonical quantization 
  scheme // J. Nonlin. Math. Phys. 2011. V. 18. 65. 
  \eprint{[arXiv:1003.1439 v.2 [hep-th]]}. 
  \emph{Korenblit~S.~E., Semenov~V.V.} On fermionic tilde conjugation rules and thermal 
bosonization. Hot and cold thermofields, // Phys. Part. and Nucl. Lett. 2011. V. 8. N 7. 
779 \eprint{[arXiv:1108.5392 [hep-th]]}.
  \emph{Semenov~V.V., Korenblit~S.~E.} Finite temperature Thirring model: from 
linearization through canonical transformations to correct normal form of thermofield 
solution \eprint{[arXiv:1109.2278 [hep-th]]} 

\bibitem {klaib} 
   \emph{Klaiber~B.} // ``Lectures in Theoretical Physics'', University of 
   Colorado (Boulder, 1967), edited by A. Barut and W. Brittin, Gordon and Breach, 
   New York, 1968, V. X, part A. 141-176.

\bibitem {d_f_z}
   \emph{Dell'Antonio~G.~F., Frishman~Y., Zwanziger~D.} Thirring model in terms of
  currents: solution and light-cone expansions // Phys. Rev.  D 6 1972. 988.

\bibitem {fab-iva}
  \emph{Faber~M., Ivanov~A.~N.} On the equivalence between Sine-Gordon model and
  Thirring model in the chirally broken phase of the Thirring model //
  Eur. Phys. J. C 20 2001. 723.  
  \emph{Faber~M., Ivanov~A.~N.} On Free Massless (Pseudo-) Scalar Quantum Field
  Theory in (1+1)-Dimensional Space-Time // Eur.Phys.J.  C 24 2002. 653. \\
  \emph{Fujita~T., Hiramoto~M., Homma~T., Takahashi~H.} New vacuum of Bethe
  ansatz solutions in Thirring model // J. Phys. Soc. Jap. 2005. V. 74. 1143. 
  \emph{Fujita~T., Hiramoto~M., Takahashi~H.} Re-interpretation of spontaneous symmetry 
  breaking in quantum field theory and Goldstone theorem. \eprint{[hep-th/0510151]}
  
\bibitem {man}
  \emph{Mandelstam S.} Soliton operators for the quantized sine-Gordon equation //
   Phys. Rev. D 11 1975. 3026. \\
   \emph{Coleman~S.} Quantum sine-Gordon equation as the massive Thirring model //
  Phys. Rev. D 11 1975. 2088.

\bibitem {blot} 
  \emph{Bogoliubov~N.~N., Logunov~A.~A., Oksak~A.~I., Todorov~I.~T.} 
  General principles of quantum field theory (Kluwer Academic Publishers, Boston, 1990)

\bibitem {oksak} 
  \emph{Oksak~A.~I.} Non-Fock linear boson systems and their applications in
  two-dimensional models // Teoret. Mat. Fiz.  1981. V. 48. 297.

\bibitem {mps_2}
   \emph{Morchio~G., Pierotti~D., Strocchi~F.} Infrared and Vacuum Structure in
  Two-Dimensional Local Quantum Field Theory Models. Fermion Bosonization // 
  J. Math. Phys. 1992. V. 33. 777.

\bibitem {i_z} 
    \emph{Itzykson~C., Zuber~J.-B.} Quantum field theory, vol. 1,2  
   (McGraw-Hill Inc., NY, 1980) 

\bibitem {rub} 
   \emph{Rubakov~V.~A.} Classical gauge fields. Theories with fermions. Noncommutative 
   theories. (URSS, Moscow, 2005)

\bibitem {raja}
  \emph{Chang~S.~J., Rajaraman~R.,} Chiral vertex operators in off-conformal theory: the 
  sine-Gordon example // Phys. Rev.  D53 1996. 2102.
}
\end{thebibliography}
\end{document}